# Charge Density Wave Modulation in Superconducting BaPbO$_3$/BaBiO$_3$ Superlattices

D. T. Harris,[1] N. Cambell,[2] C. Di,[3] J.-M. Park,[3] L. Luo,[3] H. Zhou,[4] G.-Y. Kim,[5] K. Song,[5] S.-Y. Choi,[5] J. Wang,[3] M. S. Rzchowski,[2] and C. B. Eom[1*]

[1]Department of Materials Science and Engineering, University of Wisconsin-Madison, Madison, Wisconsin 53706, USA.
[2]Department of Physics, University of Wisconsin-Madison, Madison, Wisconsin 53706, USA.
[3] Department of Physics and Astronomy and Ames Laboratory-U.S. DOE, Iowa State University, Ames, Iowa 50011, USA
[4]Advanced Photon Source, Argonne National Laboratory, Argonne, Illinois 60439, USA
[5]Department of Materials Science and Engineering, POSTECH, Pohang 37673, South Korea.

The isotropic, non-magnetic doped BaBiO$_3$ superconductors maintain some similarities to high-$T_c$ cuprates, while also providing a cleaner system for isolating charge density wave (CDW) physics that commonly competes with superconductivity. Artificial layered superlattices offer the possibility of engineering the interaction between superconductivity and CDW. Here we stabilize a low temperature, fluctuating short range CDW order by using artificially layered epitaxial (BaPbO$_3$)$_{3m}$/(BaBiO$_3$)$_m$ (m = 1-10 unit cells) superlattices that is not present in the optimally doped BaPb$_{0.75}$Bi$_{0.25}$O$_3$ alloy with the same overall chemical formula. Charge transfer from BaBiO$_3$ to BaPbO$_3$ effectively dopes the former and suppresses the long range CDW, however as the short range CDW fluctuations strengthens at low temperatures charge appears to localize and superconductivity is weakened. The monolayer structural control demonstrated here provides compelling implications to access controllable, local density-wave orders absent in bulk alloys and manipulate phase competition in unconventional superconductors.

*Correspondence to: eom@engr.wisc.edu

The properties of quantum materials are often determined by multiple distinct order parameters that represent the importance of coupled and competing phases. The high-$T_c$ cuprates exhibit a rich phase diagram with magnetic and electronic phases that are commonly thought to compete with the emergence of superconductivity [1]. In addition to magnetic ordering, recently discovered striped charge ordering in cuprates appears to compete with superconductivity in a manner reminiscent of the emergent charge density wave (CDW) seen in other superconductors [2]. Doped BaBiO$_3$ (BBO) superconductors possess a commensurate CDW in the parent compound, stable to high temperatures, in an isotropic, nonmagnetic complex oxide. The CDW manifests as bond disproportionation through a frozen breathing mode of the BiO$_6$ octahedra, as pictured in Figure 1(a), with recent experimental and first principle work suggesting hybridization of the BiO$_6$ octahedra plays a dominate role [3–6]. Doping BBO with Pb in BaPb$_{1-x}$Bi$_x$O$_3$ (BPBO, $T_c$=11 K) or K in Ba$_{1-x}$K$_x$BiO$_3$ (BKBO, $T_c$=30 K) destroys the long range CDW [7–9]. However, Raman and terahertz measurements of bulk single crystals suggest short range CDW fluctuations persist on the Bi-rich, underdoped side of the superconducting dome of BPBO and could influence both the normal state and superconducting properties [10,11].

The CDW and monoclinic distortion in BBO are similar to those found in the rare-earth nickelates [12,13], a class of materials heavily explored for possible high-$T_c$ superconductivity analogous to the cuprates and predicted to display high-$T_c$ superconductivity when grown in artificial superlattices [14]. While these non-cuprate superconductors have not been demonstrated, artificial superlattices of complex oxides can stabilize non-equilibrium phases that exhibit enhanced properties or novel phases for a wide variety of functional systems not accessible by their bulk counterparts [15]. For example, manganite-cuprate superlattices manipulate the charge-ordering phase and control electron-phonon coupling between layers [16,17]. Similarly, here we successfully stabilize, a short-range CDW order in artificial (BaPbO$_3$)$_{3m}$/(BaBiO$_3$)$_m$ superconducting superlattices with the same overall composition as the optimally doped BaPb$_{0.75}$Bi$_{0.25}$O$_3$ alloy that otherwise does not exhibit this same order. The superconducting samples show a remarkably strong temperature and superlattice period dependent CDW signature in Raman measurements due in part to charge transfer between BBO and BaPbO$_3$ (BPO) layers. This is distinct from conventional Raman modes in the BPBO alloy that show a very weak temperature dependence [10, 20]. The enhancement of the short range CDW fluctuations at low temperatures, exclusively in superlattices, is shown to increase the tendency of charge localization and weaken the superconductivity. The tuning method demonstrated here by monolayer engineering provides a platform that the competing electronic order and its fluctuations at small scales can be tailored for both elucidating the mechanisms of unconventional superconductivity and access short range emergent orders. The compelling implications may impact the entire field of emergent materials phase discovery and control.

We grew $BPO_{3m}/BBO_m$ superlattices using 90° off-axis magnetron sputtering, with the BPO:BBO fixed at 3:1, the optimal doping in the random alloy system. The total sample thickness is fixed at ~85 nm to avoid thickness related disorder effects when characterizing superconductivity [18]. We characterized film quality using laboratory and synchrotron x-ray diffraction (XRD), atomic force microscopy (AFM), and scanning transmission electron microscopy (STEM) (see supplemental information for additional growth and characterization details). Figure 1 shows characterization of our superlattice quality. Out-of-plane 2θ- θ scans exhibit a single film peak and clear Kiessig fringes originating from the uniformity of the overall film thickness. All superlattices show additional satellite reflections arising from the artificial periodicity in our grown samples, and these superlattice reflections and thickness fringes are well matched by simulations of our film (Fig. 1 (c)), which underscore the high structural quality of the superlattice samples. The out-of-plane rocking curve shows a full width at half maximum of 0.020°, indicative of the high of crystalline quality of our samples. AFM, Fig. 1(d), reveals a smooth surface with RMS of 0.5 nm and a step-terrace structure. Reciprocal space maps of the $103_{pc}$ reflection reveal relaxation of the film with respect to the substrate, expected due to the large lattice constants of BBO (~4.36 Å) and BPO (~4.25 Å), giving on the order of 10% lattice mismatch with $SrTiO_3$.

Due to the extreme electron beam sensitivity of BBO materials, atomic-resolution transmission electron microscopy quickly degraded the films making reliable chemical mapping challenging (see supplemental information and ref. [19]). We therefore performed STEM characterization of $BaBiO_3/BaPbO_3$ together with crystal truncation rod (CTR)/coherent Bragg rod analysis (COBRA) to resolve the interfacial width between the BBO and BPO layers. Low-resolution TEM, shown in Fig 2(a), reveals well-separated layers, however the similarity of the two constituent materials limits atomic number contrast between the two layers. For this reason, we carried out CTR measurements slightly beneath the Pb $L_3$ absorption edge to enhance the atomic-form-factor contrast between the Bi and Pb elements (supplementary Figure 1). The CTR measurements were performed on thinner BPO/BBO/BPO trilayer samples to allow subsequent COBRA analysis to determine the interfacial width. In Figs. 2(b) and 2(c), the COBRA-derived total electron density profile and integrated electron number in the B-site atomic layer exhibit a relatively sharp BPO/BBO interface (less than 2 unit cells in width) suggesting a limited intermixing of Pb and Bi across the interface, which further confirms the superlattice quality that gives rise to satellite peaks and places limits on the amount of interdiffusion that could lead to extrinsic effects.

Figs. 3(a) and 3(b) plot temperature dependence of carrier densities extracted from Hall measurements and transition temperatures extracted from resistivity measurements, respectively. Specifically, superlattices with small period ($m$=1,2,3) show zero resistivity and onset of a diamagnetic signal at low temperature, as shown in Figs. 3(c) and 3(d), confirming superconductivity in our superlattices. The resistive transitions are broad and multistep, possible evidence of a granular nature to the superconductivity. Samples with larger period ($m$=10, 17) do not show superconductivity above 2K, the temperature dependent resistivity and Hall carrier density remain nearly constant from room temperature to 2 K, and the behavior and the values for $n_{3D}$ closely match a reference BPO film, indicating large period films with low interfacial density behave as independent layers. In contrast, as the lattice modulation period $m$ decreases, the temperature dependent behavior of the superlattices reveals distinct differences from the bulk alloyed reference sample and possible indication of trapping of mobile charge. This trapping is seen in the more negative $d\rho/dT$ and drop in Hall $n_{3D}$ at 40 K for $m$=1,2,3 samples, and could be related to strengthening of a short range CDW that is discussed below. The room temperature $n_{3D}$ increases smoothly as $m$ is decreased and the number of interfaces increases, consistent with approximately half an electron per formula unit transferred at each interface. This means that the electron concentration in the BaBiO3 layer increases as the period of the superlattice decreases.

Further evidence for charge transfer between the layers can be found in optical reflectivity spectra shown in Figs. 4(a) and 4(b). For a BBO control sample, a strong maximum appears in optical reflectivity spectra which is a characteristic of a well-defined oscillator from the CDW mode. For $m$=2 and 5, the optical reflectivity spectra with minima around 2 eV and sharp rise in the low-energy region indicate a low-frequency spectral weight in the form of a carrier plasma.

This indicate that the charge transfer occurring in *m*=2 and m=5 films effectively dopes the BBO layer and suppress the long range CDW order. A model calculation consisting of the dielectric function of the CDW gap for BBO and a Drude carrier plasma term for the metallic film on substrate (supplementary) reproduces the experimentally determined optical reflectivity, as shown in Fig. 4(b). Lowering layer thickness *m* increases charge transfer between layers, i.e., the blueshift of the plasma edge in Fig. 4(a), possibly due to interlayer proximity and coupling, leading to higher carrier density, fully in agreement with the Hall measurements (Fig. 3(a).

To understand how charge transfer modulates the CDW, we next present temperature and superlattice period dependent Raman scattering measurements which provide evidence for the formation of a short range CDW order competing with superconductivity. The Raman spectra from $BPO_{3m}/BBO_m$ superlattices are presented in Figs. 4(c) and 4(d) for different *m*=1, 2, 3, and 10 together with control samples from BPBO and BBO films, and a STO substrate at 4.2K and 785 nm(Fig. 4(c)),  A fully gapped, long-range CDW state in BBO (gray line) gives rise to a sharp Raman active phonon breathing mode at 569 $cm^{-1}$ [19], which in the *m*=10 superlattice shows a smaller intensity and broader linewidth. Further decreases in *m* completely suppress the 569 $cm^{-1}$ peak at room temperature, yet, most intriguingly, the spectral weight of the CDW appears to persist at low temperatures as a broad spectral resonance near 550 $cm^{-1}$, even for samples showing superconductivity, e.g., as seen in *m*=3 (black line) and 2 (red line) traces. We emphasize three key properties of the "residual" spectral mode seen in the Raman spectra:

(1) It exhibits remarkably strong temperature dependence. The *m*=3 sample (black line) in Fig. 4(d) shows a featureless spectral profile near 550 $cm^{-1}$ close to room temperature that resembles the STO substrate (orange line). By removing the scattering contribution from the STO substrate in the measured spectral range of 430-680 $cm^{-1}$, normalized *I(T)/I*(300K) spectra, shown Fig. 4(e) and its top panel, clearly demonstrate the strong suppression of the residual Raman mode at elevated temperatures. Several broad resonances above 200 $cm^{-1}$ were seen and assigned as two-phonon scattering modes in the bulk single-crystal BPBO [20]. However, these modes only exhibit a very weak temperature dependence with large scattering intensity at room temperature which cannot account for the strikingly sensitive temperature tuning of the presently observed residual Raman mode.

(2) The broad, Raman spectral shape in Fig. 4(e) is characteristic of two resonant modes at 598 cm$^{-1}$ and 545 cm$^{-1}$, marked by two dashed lines. The 598 cm$^{-1}$ peak has been assigned to the breathing mode of BiO$_6$ octahedra [19], which has the same origin as the 569 cm$^{-1}$ CDW mode in BBO in Pb-doped metallic compounds. The energy increase and spectral lineshape change are likely due to the decrease of the Bi-O bond length [20] and/or the electron-phonon renormalization. Therefore, in the absence of the sharp CDW peak from the long-range order, these distinct monolayer number $m$ and temperature dependences of the residual resonances provide a compelling evidence for the existence of short-range CDW correlations for samples showing superconductivity. This local electronic order strongly renormalizes the breathing mode that gives rise to the persisting spectral weight with a strong temperature dependence absent in bulk single crystal BPBO [19, 20].

(3) The strong $m$ dependence of the residual spectral weight further demonstrates the monolayer thickness control of CDW fluctuations in our superlattice systems. This is clearly seen in Fig. 4(c) where the residual Raman mode for $m=3$ (black) is significantly larger than both $m=1$ (blue) and BPBO bulk alloy (green) samples. Prior studies of bulk single crystals related optical spectra and pseudogap formation with a spatially-temporally fluctuating short-range CDW order of a few unit cells in BiO$_6$ octahedra on the Bi-rich, underdoped side of the BPBO phase diagram [10, 11], which is consistent with our proposed physical mechanism here. It is critical note, however, that such short range CDW is not present in the optimally doped BaPb$_{0.75}$Bi$_{0.25}$O$_3$ alloy with the same overall chemical formula as our superlattice samples.

This work demonstrates structural and electronic engineering of a model system BPO$_{3m}$/BBO$_m$ at the monolayer limit for understanding CDW that competes with superconductivity. The design and discovery of previously-unestablished superlattices allow the manipulation of the CDW and its fluctuations by controlling charge transfer. This hints additional quantum control of hidden phases by combining with other complementary tuning methods [24, 25].

**Acknowledgements**

Synthesis, characterization, and analysis at UW-Madison were supported with funding from the Department of Energy Office of Basic Energy Sciences under award number DE-FG02-06ER46327. The work at Ames Laboratory and Iowa State University (Raman and optical spectroscopy characterizations and analysis) was supported by the DOE Office of Science, Basic Energy Sciences, Materials Science and Engineering Division. Ames Laboratory is operated for the DOE by Iowa State University under Contract No. DE-AC02-07CH11358. This research used resources of the Advanced Photon Source, a U.S. Department of Energy (DOE) Office of Science User Facility operated for the DOE Office of Science by Argonne National Laboratory under Contract No. DE-AC02-06CH11357.

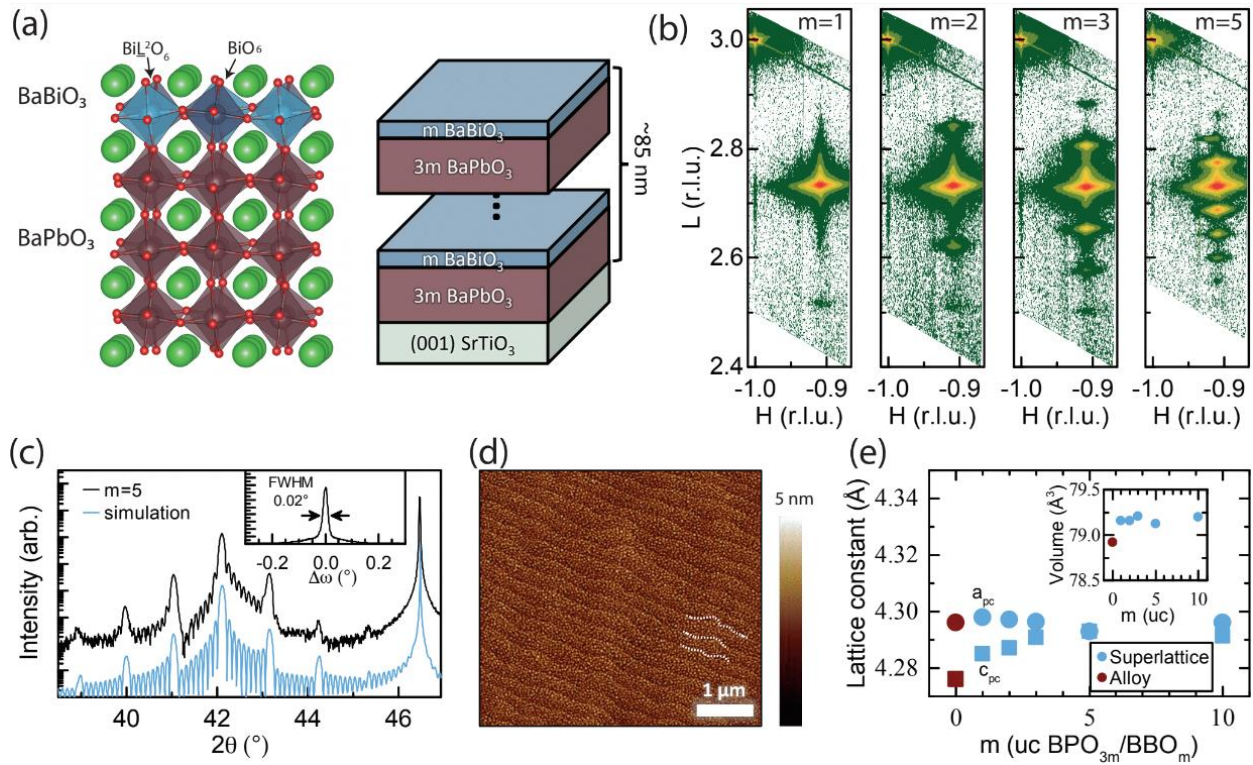

**Figure 1.** Superlattice characterization. (a) Schematic of $BaBiO_3$, $BaPbO_3$, and superlattice structures. (b) Reciprocal space maps around the $103_{pc}$ reflection for m=1, 2, 3, and 5 superlattices. Satellite superlattice reflections are visible for all samples. (c) Out-of-plane $2\theta$-$\theta$ scan of an m=5 sample showing clear Kiessig fringes and superlattice reflections with excellent agreement with the pattern simulation. The inset shows the out-of-plane rocking curve of the film reflection. (d) Atomic force microscopy image of the surface topography. The white dashed lines highlight step edges. (e) In-plane and out-of-plane pseudocubic lattice parameters extracted from reciprocal space maps. The inset shows the unit cell volume.

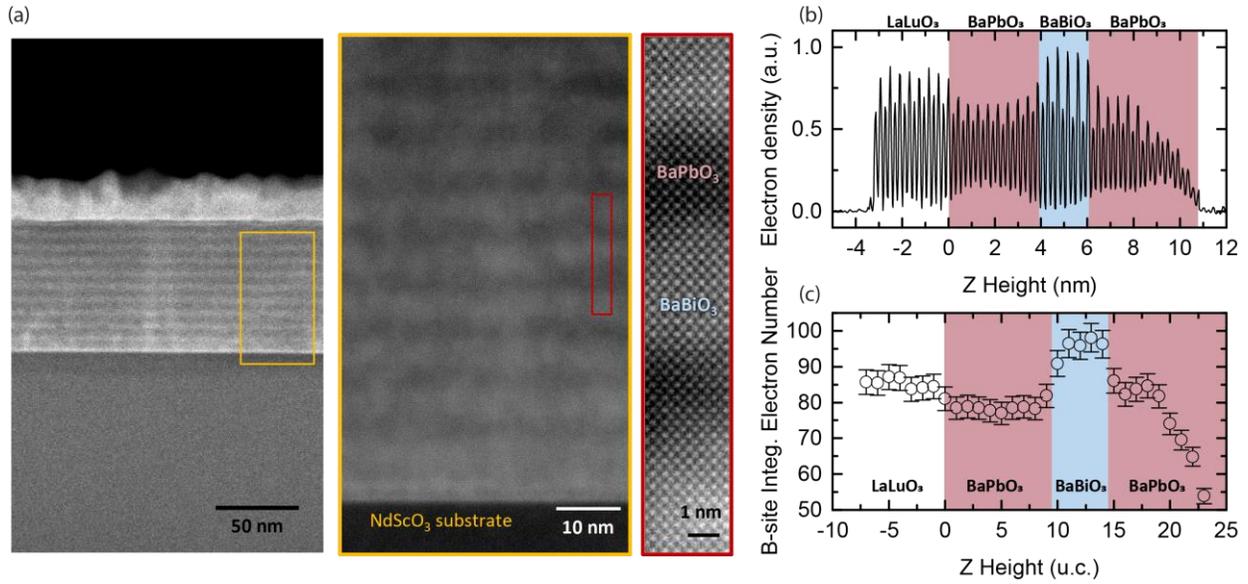

**Figure 2.** TEM characterization of BaBiO$_3$/BaPbO$_3$ and chemical interfacial width and intermixing at the interfaces resolved by the resonant coherent Bragg rod analysis method. (a) Low and high resolution STEM images. (b) Resonant COBRA-derived electron density profile of a BaPbO$_3$/BaBiO$_3$/BaPbO$_3$ trilayer sample. (c) Resonant COBRA-derived integrated electron number in each perovskite B-site atomic layer along the out-of-plane direction.

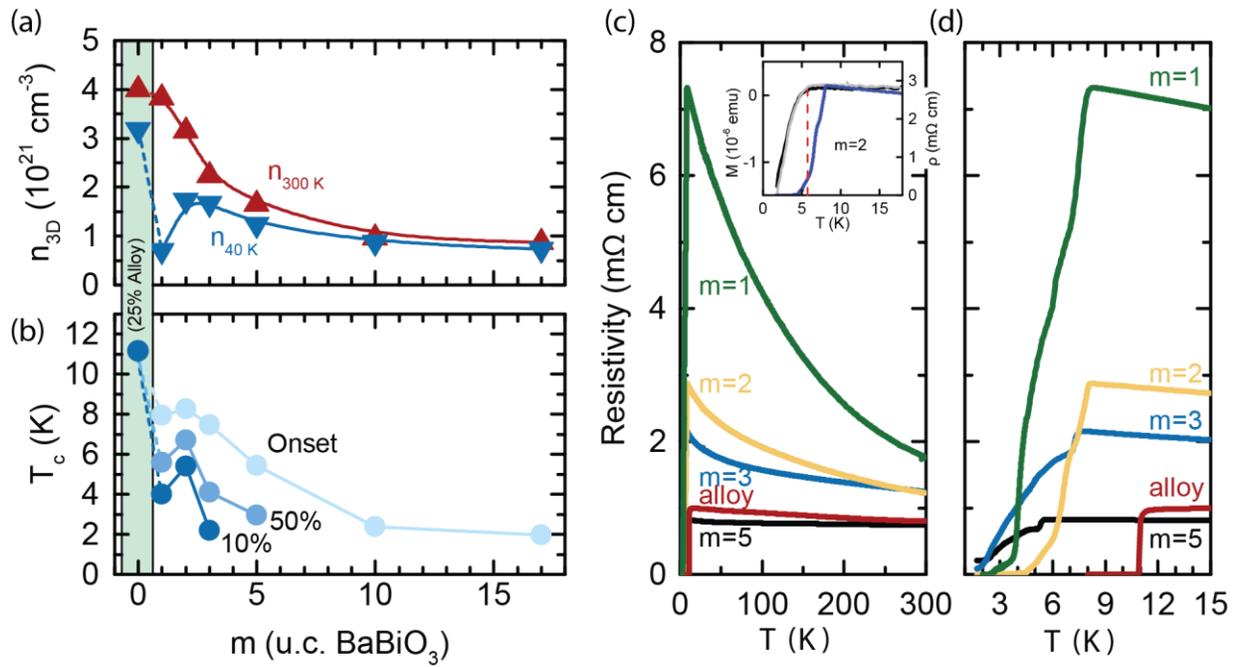

**Figure 3.** Transport characterization. (a) Room and low temperature 3D carrier densities extracted from Hall measurements. (b) Transition temperatures extracted from resistivity measurements. (c) Full temperature range resistivity vs. temperature. The inset of (c) overlays the transition probed by resistive measurements (blue) with out-of-plane (black) and in-plane (grey) onset of diamagnetism (red dashed line) for *m*=2. (d) Transition region of superconducting samples.

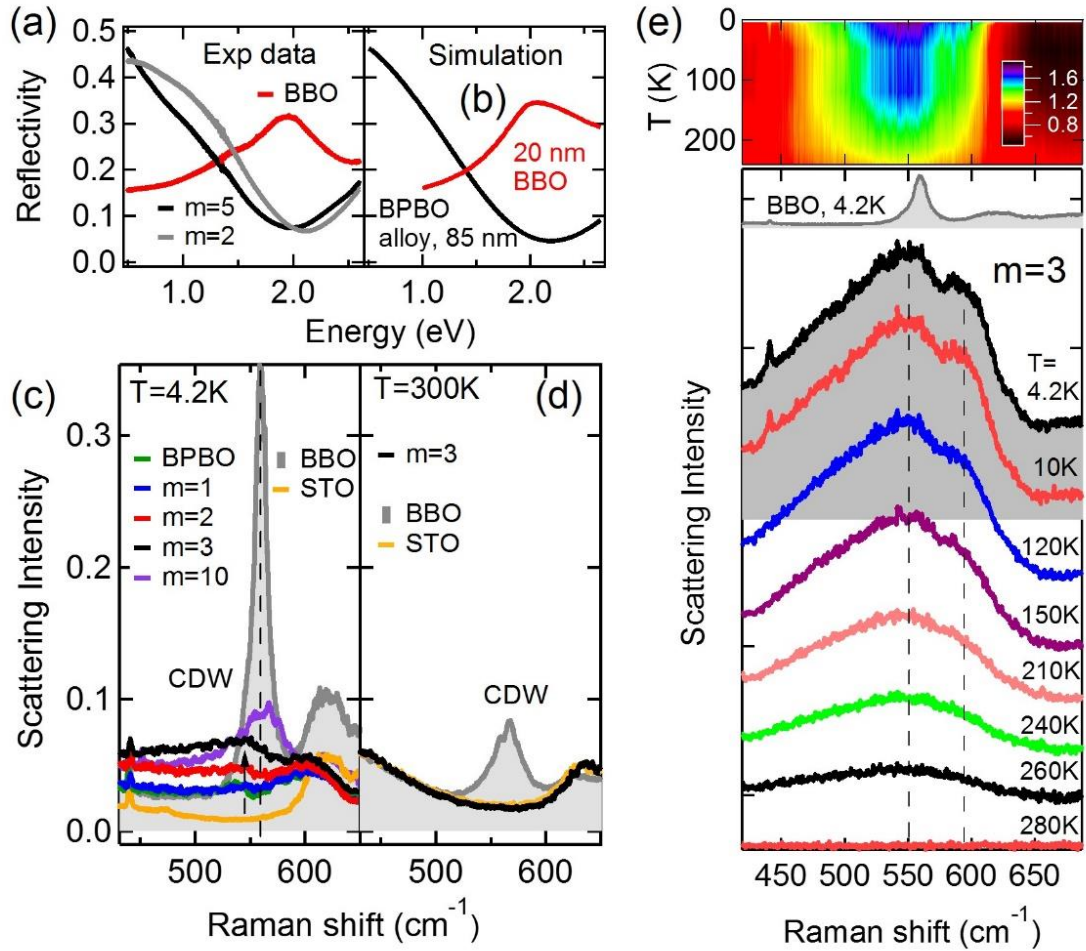

**Figure 4.** Optical and Raman spectroscopy characterizations. (a) Experimental reflectivity data for BBO and for *m*=2 and 5 superlattices. (b) Simulation of the optical spectra [21-23].. (c) Low temperature Raman measurements around the CDW peak (dashed line) for selected superlattices, reference BBO and BPBO films, and STO substrate. The charge transfer (main text) suppresses the sharp CDW peak to a broad cusp shape of spectrum (black arrow). (d) Room temperature Raman measurements in an *m*=3 superlattice. (e) Normalized low temperature Raman spectra (*I*(*T*)/*I*(300K)) around the CDW peaks for the *m*=3 sample. The traces are offset for clarity.

Supplementary Materials for

# Charge Density Wave Modulation in Superconducting BaPbO$_3$/BaBiO$_3$ Superlattices

D. T. Harris,[1] N. Cambell,[2] C. Di,[3] J.-M. Park,[3] L. Luo,[3] H. Zhou,[4] G.-Y. Kim,[5] K. Song,[5] S.-Y. Choi,[5] J. Wang,[3] M. S. Rzchowski,[2] and C. B. Eom[1*]

[1]Department of Materials Science and Engineering, University of Wisconsin-Madison, Madison, Wisconsin 53706, USA.
[2]Department of Physics, University of Wisconsin-Madison, Madison, Wisconsin 53706, USA.
[3] Department of Physics and Astronomy and Ames Laboratory-U.S. DOE, Iowa State University, Ames, Iowa 50011, USA
[4]Advanced Photon Source, Argonne National Laboratory, Argonne, Illinois 60439, USA
[5]Department of Materials Science and Engineering, POSTECH, Pohang 37673, South Korea.

*Correspondence to: eom@engr.wisc.edu

***Thin Film Growth:*** Thin films were grown from BaBiO$_3$ and BaPbO$_3$ targets containing 13% excess Bi or Pb in the initial powders to account for volatility during processing and growth. Films were grown using 90° off-axis rf-magnetron sputtering with computer-controlled, pneumatically actuated shutters to precisely control the layer thicknesses. Growth rates were calibrated by measuring the growth rate using x-ray reflectivity prior to superlattice growth. Complete details of the target and film preparation are given in the supplemental information of [18].

***Superconducting characterization:*** Magnetotransport measurements were performed using sputtered Ag/Au electrodes in a Quantum Design Physical Properties Measurement System helium crystostat from 1.8 K to 300 K in a Van der Pauw geometry and the total superlattice thickness. Magnetic susceptibility measurements were performed in a Quantum Design Magnetic Properties Measurement System from 1.8K to 300K with applied in-plane and out-of-plane fields of magnitude less than 100Oe.

***Optical reflectivity*:** The reflectivity was measured with Spectrometer >1.3eV and FTIR <1.3eV with Al reference. The measured reflectivity was corrected with Al reflectivity. Three phase (air/BBO (or BPBO)/STO) thin film model was used to simulate reflectivity of different thickness BBO thin films [22]. Optical constants were obtained by a single oscillator fitting from BBO spectrum [23]. STO $n$, $k$ were used from book value [21]. By using the corresponding film thickness for superlattice samples, such a model provides good agreement with the experimental behavior with salient features from the CDW peak (red line) and the plasma edge (black and gray lines for $m=5$ and $m=2$ in Fig. 4(a), respectively). Note that the plasma edge changes in Figs. 4(a)-4(b) corroborate the reduction of the residual spectral weight associated with the short-range CDW order in smaller period samples, as discussed in Fig. 4(c)-4(e). This is consistent with the fact that the amount of charge transfer increases with decreasing superlattice number $m$ shown in the optical reflectivity measurement in Fig. 4(a). This indicates that the monolayer control demonstrated provides a new tuning knob for controlling density-wave fluctuations and phase competition beyond bulk alloys.

***Raman spectroscopy*:** Raman spectra were acquired using three Notch filters (BNF) in combination with a single monochromator [26] equipped with a CCD detector. All measurements were performed with samples placed in a liquid-helium-cooled cryostat in backscattering configuration using a 784.5 nm continuous wave diode laser. The laser was cleaned up by four band pass filters (BPF) and the excitation power on the sample was ~4 mw. The acquisition time was 150s for all measurements.

***Crystal truncation rod (CTR) measurement and coherent Bragg rod analysis (COBRA)*:**
The CTR measurement of the $BaPbO_3/BaBiO_3/BaPbO_3$ trilayer sample was performed on a five-circle diffractometer with χ-circle geometry at sector 12-ID-D of the Advanced Photon Source, Argonne National Laboratory. The X-ray energy was tuned to just below the Pb $L_3$ absorption edge (E = 13.032 keV) to enhance the contrast of the atomic form factors between Bi and Pb elements. The X-ray beam at the beamline 12-ID-D has a total flux of $5 \times 10^{12}$ photons/s and was vertically focused by beryllium compound refractive lenses down to a beam profile of ~ 50 μm at this energy. In the COBRA analysis to make convergence on satisfactory electron density profile, unbiased starting structural parameters with all bulk-like settings were selected for subsequent iterations. The experimentally measured and COBRA simulated specular CTR (00L) are shown in Figure 5.

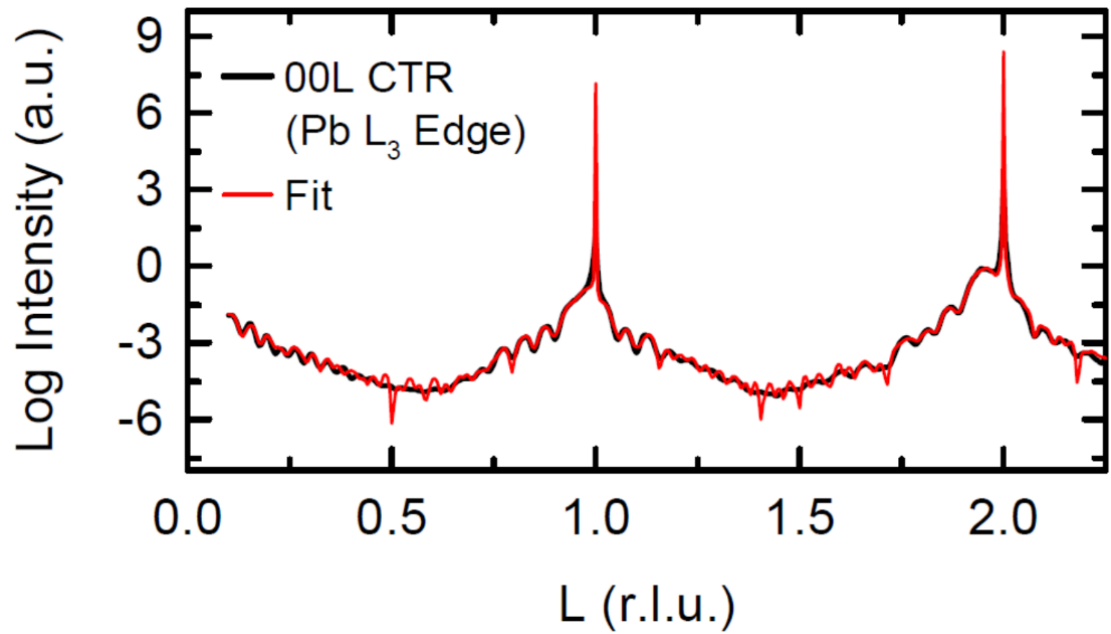

**Figure 5.** Experimentally measured and the COBRA simulated specular (00L) crystal truncation rod with the X-ray energy tuned to just below the Pb $L_3$ absorption edge.

***Scanning transmission electron microscopy (STEM):*** The sample for STEM analysis was prepared through mechanical flat polishing by using a precise polishing system (EM TXP, Leica), and gently polishing down to a thickness of ~10 μm. To avoid the decomposition of BBO/BPO by water, we used the water-free lubricant (Purple lubricant, Allied). The polished specimen was ion-milled using Ar ion beam with voltage range of 1-3 keV (PIPS II, Gatan) to make the hole for the STEM observation. To reduce damage during ion-milling, we used a cold stage fixed at -100 °C to limit the fracture from thermal shock of the $NdScO_3$ substrate. After making an electron-transparent area, low energy milling was performed using 0.1 keV Ar beam to minimize surface damage from the prior ion-milling process.

   To avoid electron beam damage during the STEM observation, STEM analysis was performed using a 30-200kV STEM (JEM-ARM200F, JEOL) at 60 kV equipped with an aberration corrector (ASCOR, CEOS GmbH). The optimum size of the electron probe was ~ 0.1 nm. The probe used for acquiring the HAADF and ABF images was 9C (23.2 pA). The collection semi-angles of the HAADF detector were adjusted from 68 to 280 mrad in order to collect a large-angle elastic scattering electrons for clear Z-sensitive images. The ABF images were obtained using a 3 mm aperture and a collection angle from 12 to 24 mrad was used. The obtained raw images were processed with a band-pass Wiener filter with a local window to reduce a background noise (HREM research Inc.). STEM image simulation was performed using Dr. Probe (free software, Ernst-Ruska Institute at Juliech). Figure 6 discusses observed damage during imaging in detail.

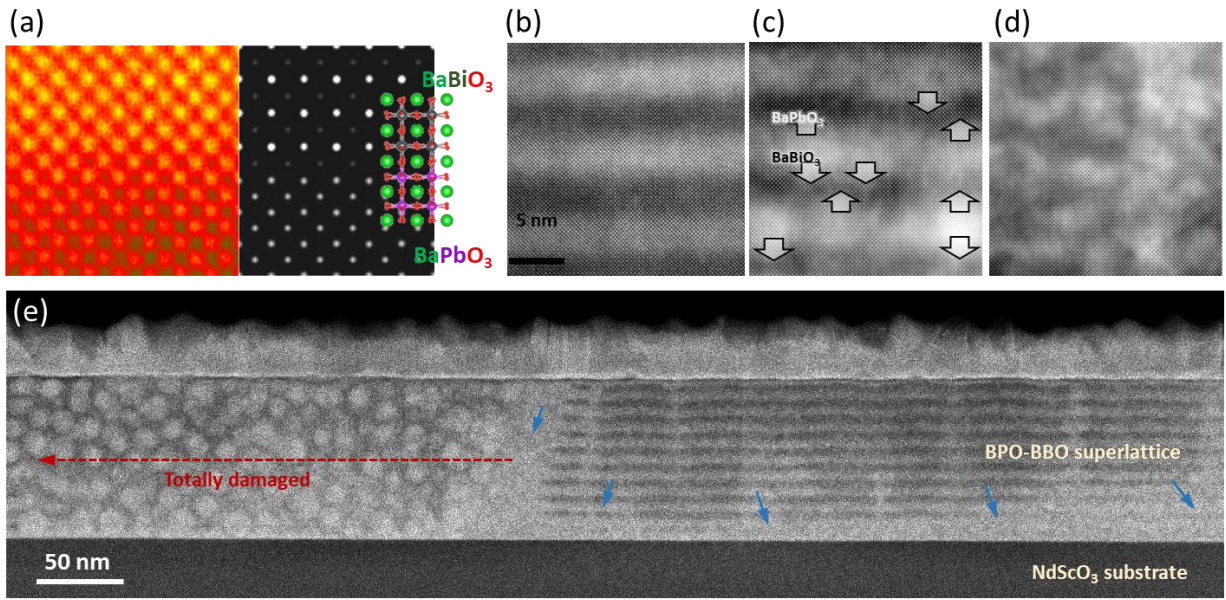

**Figure 6.** Damage during STEM imaging. (a) The interface between BaBiO$_3$ and BaPbO$_3$ is shown. Despite the insignificant difference in the atomic weights between Bi and Pb, the interface is visible with a brighter BaBiO$_3$ layer and a darker BaPbO$_3$ layer. This contrast difference seems to arise from crystallographic differences; the octahedral tilt in the less symmetric, monoclinic BaBiO$_3$ gives rise to stronger dechanneling effect of the electron probe. The image on the right side is the simulated STEM HAADF image of the interface structure of monoclinic BaBiO$_3$ and orthorhombic BaPbO$_3$. The brighter contrast in BaBiO$_3$ is found using Dr probe simulator (Ernst-Ruska Institute at Juliech). However, it should be noticed that other simulators such as JEMS (Interdisciplinary center for electron microscopy) and QSTEM (Humboldt University) could not realize the brighter contrast in BaBiO$_3$. This unexpected contrast difference between BaBiO$_3$ and BaPbO$_3$ requires the further investigation. (b), (c) and (d) images are the time-sequential series of HAADF STEM images showing the BaBiO$_3$ and BaPbO$_3$ film being damaged by electron beam during the STEM observation even under 60kV of the acceleration voltage. Arrows in (c) indicate that the blurred, white contrast related with decomposition starts to grow from the BaBiO$_3$ to the BaPbO$_3$ layer. Later on, the BaBiO$_3$ - BaPbO$_3$ superlattice film is fully damaged as shown in (d). (e) Some regions in the sample was already damaged possibly by the Ar ion beam during ion-milling. The left side of (e) shows a totally damaged film so that the BaBiO$_3$ - BaPbO$_3$ superlattice cannot be observed at all. The right side of (e) shows the superlattice having the periodic white layers (BaBiO$_3$) and black layers (BaPbO$_3$) still survive whilst some areas are degraded to the amorphous phases by the electron beam, indicated by the blue arrows. The atomic scale observation of BaBiO$_3$ - BaPbO$_3$ superlattice film in this study was only available at the somewhat thick area and also the short exposure was essentially required to minimize the exposure time to the electron beam.